# DESIGN AND HARDWARE IMPLEMENTATION OF A SEPARABLE IMAGE STEGANOGRAPHIC SCHEME USING PUBLIC-KEY CRYPTOSYSTEM

Salah Harb[1], M. Omair Ahmad[1] and M.N.S Swamy[1]

[1]Electrical and Computer Engineering Department, Concordia University, 1440 De Maisonneuve, Montreal, Canada
`{s_rb,omair,swamy}@ece.concordia.ca`

## ABSTRACT

*In this paper, a novel and efficient hardware implementation of steganographic cryptosystem based on a public-key cryptography is proposed. Digital images are utilized as carriers of secret data between sender and receiver parties in the communication channel. The proposed public-key cryptosystem offers a separable framework that allows to embed or extract secret data and encrypt or decrypt the carrier using the public-private key pair, independently. Paillier cryptographic system is adopted to encrypt and decrypt pixels of the digital image. To achieve efficiency, a proposed efficient parallel montgomery exponentiation core is designed and implemented for performing the underlying field operations in the Paillier cryptosystem. The hardware implementation results of the proposed steganographic cryptosystem show an efficiency in terms of area (resources), performance (speed) and power consumption. Our steganographic cryptosystem represents a small footprint making it well-suited for the embedded systems and real-time processing engines in applications such as medical scanning devices, autopilot cars and drones.*

## 1. INTRODUCTION

In the modern era of technology, sharing multimedia content has become easier and faster. As a result of that, malicious tampering and unauthorized data manipulation have been more accessible to the eavesdroppers over the communication channels. To prevent that, hiding data technique is one of the possible solutions that provides a reliable, safe and secure communication channels in applications such as image authentication, copyrights, and fingerprinting [1]. Hiding data in a carrier is referred to as steganography, which is the art of hiding secret data in a carrier in such a stealthy way that avoids the suspicion of unauthorized receivers.

Secret data can be concealed in various carriers, but digital images are the most preferable and suitable carrier due to massive users and applications that have the frequent access to it on internet. Moreover, performing image filtering techniques, mechanisms and cryptographic systems over digital images is more efficient in the reconfigurable hardware platforms. There are three main key characteristics that determine the performance of steganographic cryptosystems: embedding rate, imperceptibly and robustness. Efficient cryptosystems have a balance realization between these three characteristics. The massive variety in digital image real-time embedded systems and applications makes them more vulnerable to the attacks from the hackers (e.g. steganalysis tools) [2] for malicious targets, as delivery services, sharing rides, spying and warfare. Securing the secret data (i.e. control data) and cover image (i.e. footage images) itself is one of the steganographic frameworks that is achieved to protect the stego image against the steganalysis attacks and providing privacy services. This implies to have another secure layer before or after concealing data, where a public-key cryptosystem is implemented by having different kinds of secret codes, which are so-called private and public keys. The computational complexity in such these frameworks is increased because of the time-resource consuming field operations presented in these public-key cryptosystems.

Hardware platforms presents a strong flexibility for such these high computational-complexity cryptosystems [3]. Application Specific Integrated Circuits (ASICs) and Field Programmable Gate Array (FPGA) are very popular hardware platforms for designing and implementing cryptographic and image steganographic cryptosystems. FPGAs can provide a high performance that can be achieved by ASIC platforms, and they cost much less than ASICs [4] [5]. FPGAs are reconfigurable and physically secure devices, which are preferable platforms to the researchers for testing their implementations. ASICs/FPGAs platforms have many features and computational capabilities that improve the performance such as parallelism and pipelining architectures. Few steganographic hardware implementations have been designed [6] [7] [8], which aim to increase the processing speed (throughput), less consumed area (resources), higher embedding rates, better image quality (PSNR) and robustness (security).

In [7], a hardware cryptoprocessor for privacy-preserving data mining algorithm is implemented using the paillier cryptosystem. Parallelism is applied through exploiting the independency among the modular operations such as multiplications and exponentiations. Pipelined stages are inserted among the field operations to break the long critical data paths. The cryptoprocessor is evaluated using privacy-preserving matching set intersection protocol. The authors provide a deep hardware realization of the privacy-preserving scheme using FPGA platform. As a case study, the cryptoprocessor is integrated into a privacy preserving set intersection protocol. A performance evaluation for the protocol is performed between the hardware and the software implementations. In [7], the parallelism is applied through duplicating instances of the independent modular operations resulting in using more resources. For example, the modular exponentiation operation is replicated twice for performing two exponents with different bases. Pipelining is done by utilizing DSP blocks to break up the long critical data paths, which leads to enhance the running frequency. Buffering the operations is employed to maintain the parallel execution between the elements of these operations. Extensive pipelining and buffering produce more cycles to perform modular operations. Modular exponentiation is performed using right-to-left binary algorithm, which includes repeating the modular multiplications based on the exponent. Right-to-left binary algorithm does not provide high performance when the base of the exponent is fixed all the time of processing intermediate values till the final result.

In this paper, a new high-performance public-key image steganographic framework is designed and implemented using FPGA reconfigurable hardware platform. The paillier public-key cryptosystem is used to encrypt pixels of the cover images [9]. Pixels of the cover image are encrypted by paillier cryptosystem with a public key to generate an encrypted image. After that, a stego image is generated by performing an embedding process where a secret data is concealed into the encrypted image. In a reversible way, the receiver, who has the private part (i.e. private key) of the public key, can decrypt the received loaded-encrypted stego image. The result is a loaded stego image that has the secret data. Extracting process is applied after. Reconstructing the cover image is done as a last step by sorting the decrypted pixels.

Figure 1 illustrates the general framework of the public-key image steganographic cryptosystem. Parallelism is applied to improve the computational complexity in the paillier cryptosystem, and efficient transitions are considered in the finite state machines that control each main component in the proposed cryptosystem. For validating purposes, the full image steganographic cryptosystem is implemented in these FPGA devices introduced by Xilinx [10].

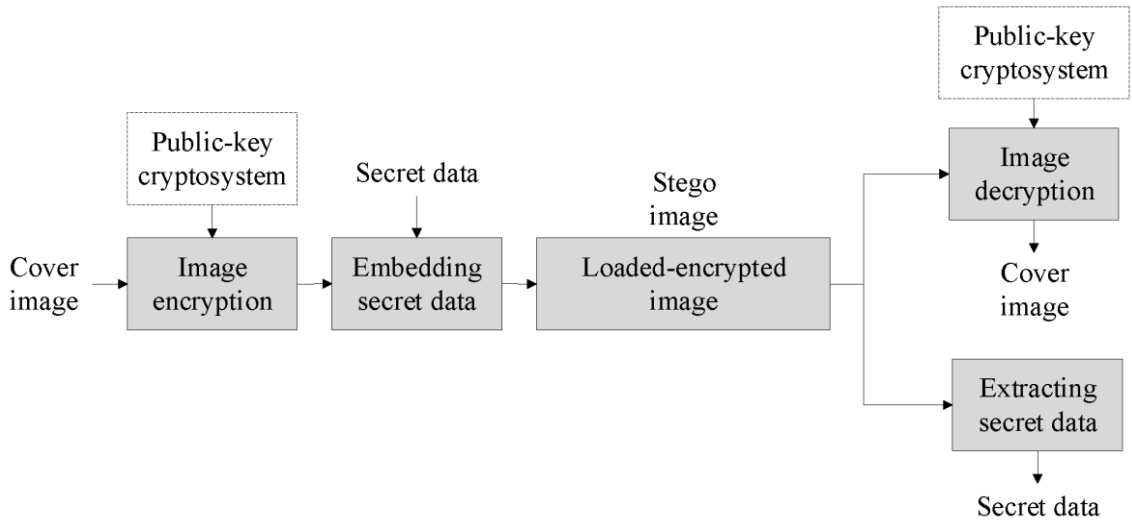

Figure 1. The general framework of the proposed public-key image steganographic cryptosystem.

This paper is organized as follows: Section 2 introduces a background for the paillier public-key cryptosystem. Section 3 presents the hardware architecture design for the proposed public-key image steganographic framework. The proposed steganographic cryptosystem evaluation in terms of embedding rate (bpp), image quality (PSNR), speed, resources utilization, power consumption and throughput are presented in Section 4. Finally, Section 5 concludes this paper.

## 2. PAILLIER CRYPTOSYSTEM

Paillier cryptosystem is a probabilistic public-key (e.g. asymmetric) cryptographic algorithm, which was invented by Pascal Paillier in 1999 [9]. It's based on the RSA computational-complexity public-key algorithm. Paillier cryptosystem has the *n-th* residue problem, where finding the composite *n-th* residue is believed to be computationally hard. The feature homomorphic property [9] in paillier cryptosystem makes it very appealing to be used and integrated in privacy-preserving embedded systems as in transferring money and electronic voting campaigns applications.

The textbook version of RSA is a deterministic public-key algorithm (i.e. no random components) [11], which makes it vulnerable to the chosen-plaintext attacks by exploiting the multiplicative property [12] [13]. In this attack, the attacker can distinguish between the ciphertexts, and this kind of RSA implementation is referred to as non-semantic cryptosystem. To resist this type of attacks, padding schemes for RSA is applied by embedding some random paddings into the plaintexts before starting encryption process [14]. This makes RSA is a semantic secure algorithm, which means the attackers could not distinguish between the ciphertexts.

Padded RSA cryptosystem does not support the homomorphic property since the randomness injected to the plaintext before encryption [15], hence RSA can be either semantically secure or homomorphic cryptosystem. On the other hand, Paillier cryptosystem is a semantic-homomorphic cryptosystem. However, the price to pay is that paillier cryptosystem consumes more resources with equivalent security level with RSA [15]. Paillier cryptosystem requires more modular exponentiation and multiplications field operations than the RSA. The next is a brief of how the encryption and decryption are processed in paillier cryptosystem.

## 2.1. Key GENERATION in Paillier Cryptosystem

At any public-key cryptosystem, there is a key generation pre-process is done by the receiver, where public and private keys are generated for the encryption and decryption processes [16]. In paillier cryptosystem, the receiver chooses two large primes, $q$ and $p$, the Great Common Divisor (GCD) for $q.p$ and $(q-1)(p-1) = 1$, A hard-to-factor number $n = q.p$ and $\lambda = LCM((q-1)(p-1))$ are obtained, where LCM is the Least Common Multiple. The receiver randomly selects $g$ in the $Z_{n^2}^*$ field, where GCD $(L(g^\lambda \bmod n^2), n) = 1$, and $L(x) = (x-1)/n$ [16]. The receiver finally distributes his public key pair $(n, g)$, and the $\lambda$ is considered as the private key for that public key.

## 2.2. Encryption and Decryption in Paillier Cryptosystem

The sender has a message $M \in Z_n^*$, then a random integer $r \in Z_n^*$ is selected for semantic security [1]. The ciphertext $C$ of that $M$ is computed using the following equation:

$$C = ENC_{pk}(M, r) = g^M r^n \bmod n^2 \qquad (1)$$

where $ENC$ is the process of encryption and $pk$ is the public key of the receiver. The ciphertext $C$ is transmitted through the communication channel, the receiver gets it, and performs the following to reveal the original message $M$:

$$M = DEC_{pr}(C) = L(C^\lambda \bmod n^2) / L(g^\lambda \bmod n^2) \bmod n \qquad (2)$$

where $DEC$ is the process of decryption, $L(x) = (x-1)/n$, and $pr$ is the private key of the receiver.

## 2.3. Homomorphic Properties in Paillier Cryptosystem

The power of paillier cryptosystem comes from the additive homomorphic properties, which makes it a very suitable cryptosystem for electronic applications that require hiding identities as privacy-preserving perspective [17]. At the encryption process, let say we have two encrypted messages, $E_1$ and $E_2$, such as: $E_1 = ENC_{pk}(M_1, r_1)$ and $E_2 = ENC_{pk}(M_2, r_2)$. These two encrypted messages are considered as an additive homomorphic function at decryption process as follows:

$$DEC_{pr}(E_1, E_2) = M_1 + M_2 \bmod n \qquad (3)$$

To verify that, check the encryption process for both messages as follows:

$$ENC_{pk}(M_1, r_1) \times ENC_{pk}(M_2, r_2) = g^{M_1} g^{M_2} r^n r^n \bmod n^2 = g^{M_1+M_2} r^n r^n \bmod n^2 \qquad (4)$$

Another form of additive homomorphic property is multiplying $ENC_{pk}(M_1, r_1)$ with $g^{M_2}$, which will give us the sum of the corresponding messages: $M_1 + M_2$. For multiplicative homomorphic property, an encrypted message $ENC_{pk}(M_1, r_1)$ is raised to a constant $k$, the decryption is the message $k.M_1$ as shown in the following equation:

$$DEC_{pr}(ENC_{pk}(M_1, r_1)^k \bmod n^2) = k.M_1 \bmod n \qquad (5)$$

## 3. THE PROPOSED HARDWARE ARCHITECTURE DESIGN

In this section, the structure of the proposed image steganographic cryptosystem is presented. The structure consists of components that works together to embed-extract and encrypt-decrypt a single pixel. We should mention that all pixels of the cover image are within gray-scale levels (e.g. 0 to 255). Each component at both encryption and decryption processes is controlled by an efficient finite state machine. Figure 2 represents the main components of the proposed steganographic cryptosystem, and paillier cryptosystem is used for encrypting and decrypting pixels.

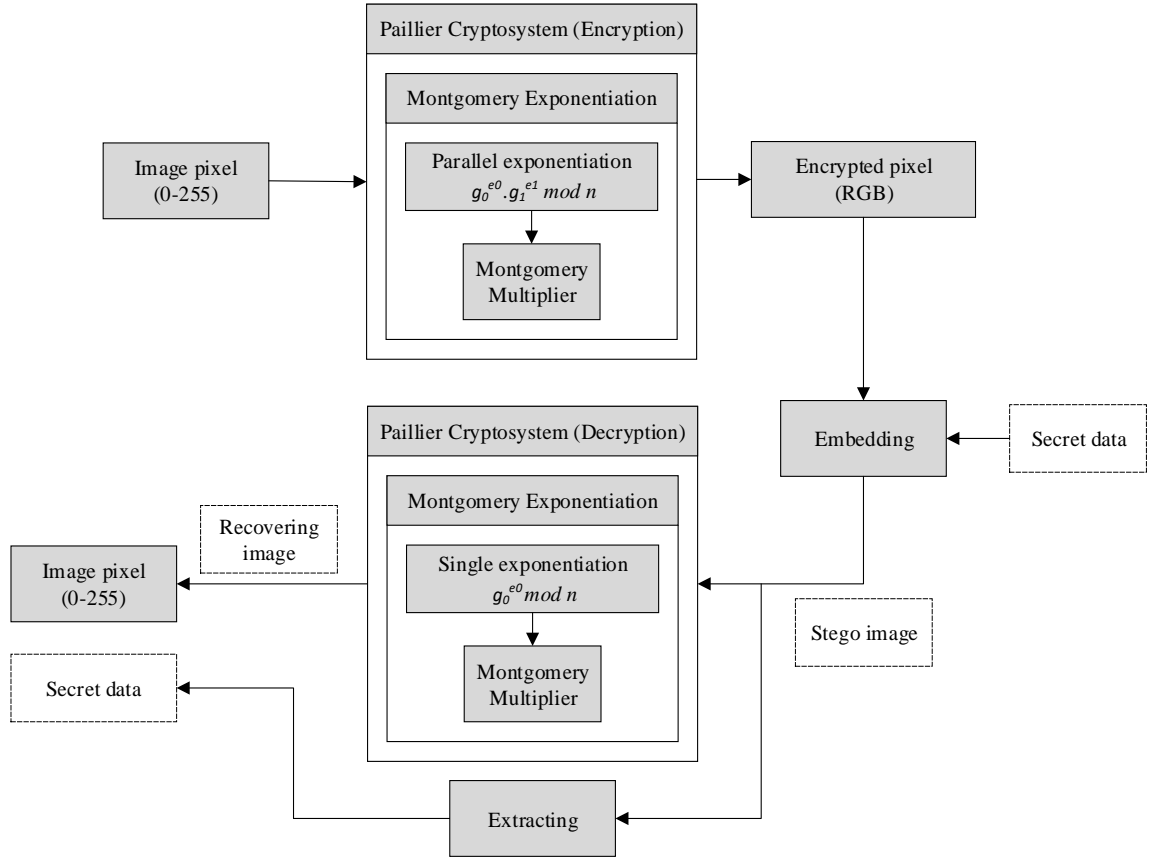

Figure 2. The main components of the proposed steganographic cryptosystem in encryption and decryption processes.

## 3.1. Image Embedding and Encryption

To protect and increase the robustness, encrypting the pixels of the cover image is performed first, and then the embedding process is done to the encrypted cover image. Least Significant Bit (LSB) is one of the low-complexity steganographic schemes that can be applied to embed the secret data into pixels of the cover image. Increasing the capacity of the concealed data (e.g. high bpp) can be achieved by embedding more bits into a single pixel. However, this results in a degradation of quality for the cover image (e.g. lower PSNR) since original bits are altered. In our proposed cryptosystem, the embedding procedure doesn't require to change any single bit in pixels of the cover image. This provides us the capability of recovering the cover image without decreasing the quality of the cover image (100% PSNR). For instance, assume the cover image has a resolution 64 rows x 64 columns (4096 pixels). the cover image can conceal up to 4096 bits of secret data, which equals to 512 bytes. Embedding single is done by using the additive homomorphic property of the paillier cryptosystem. Each pixel P is divided into two values ($M_1$ and $M_2$) in such a way that $P = M_1 + M_2$. Each value is encrypted using Equation 1 to obtain $E_{M1}$ and $E_{M2}$. The embedding is done as follows; if secret bit is 1 and $E_{M1} < E_{M2}$, Swap $E_{M1}$ and $E_{M2}$ values. If secret bit is 0 and $E_{M1} > E_{M2}$, Swap $E_{M1}$ and $E_{M2}$ values. The embedding is done when all bits of the secret data are scanned. Note that sequence of selected pixels is defined using a data hiding key shared between sender and receiver.

Equation 1 states that, to encrypt pixel (e.g. plain text), we should raise g to the power of pixel value, and multiply it with a randomly selected *r* value to the power of *n*. The result then is reduced (*mod*) to the selected field ($n^2$). These two exponentiations over modulus are very common in modular arithmetic computations in public-key cryptography. Single modular exponentiation (*C*) computes the remainder of the base (*b*) raised to the exponent (*e*) power as $C = b^e \bmod n$ where $0 \leq C < n$. Performing this kind of modular operations is done by using different algorithms. Binary Left-to-Right exponentiation algorithm is one of the easiest and trivial way to compute $g^e$. In binary algorithm, the bits of *e* are scanned, then perform 1 squaring every time and 1 field multiplication when the current bit equal to 1. In 1985, P. Montgomery proposed a new algorithm to compute the modular multiplication operations efficiently [18]. He proposed to map the presentation of the elements in any field $Z_n$ to a corresponding domain, which is called as Montgomery Domain (MD). Assume the modulus *n* and *x* are integers in $Z_n$, Consider *R* is the radix $2^k$, where *k* is the number of bits in modulus *n*, and GCD (*n*, *R*) = 1. Mapping the *x* to the montgomery domain is done as $X_{MD} = x \cdot R \bmod n$. The natural representation of the *x* in montgomery domain can be obtained by multiplying the $X_{MD}$ by the multiplicative inverse of the considered *R* radix $x = X_{MD} \cdot R^{-1} \bmod n$. Multiplying *x* and *y* is done by first moving them to the montgomery domain, then observing the $Z_{MD} = X_{MD} \cdot Y_{MD} \bmod n$ can hold the following:

$$X_{MD} \cdot Y_{MD} = x \cdot R \cdot y \cdot R = z \cdot R^2 \bmod n = Z_{MD} \cdot R \bmod n \tag{6}$$

A reduction is required to obtain the result $Z_{MD} = Z_{MD} \cdot R^{-1} \bmod n$, which a montgomery reduction is applied for that easily. Since the *z* is what we are looking for, the montgomery multiplication combines the reduction $Z_{MD} \cdot R^{-1} = z$ and multiplication $x \cdot y$ operations to compute the product of two integers. Performing the montgomery modular exponentiation $g^e \bmod n$ can be done by combining the binary exponentiation and the montgomery multiplication. Algorithm 1 shows this kind of combination. It consists of *k* executions for the main loop, which contains two montgomery multiplication. For single montgomery multiplication, the computational time is to $3 \cdot k \cdot T$ [19], where T is the execution time for a single full-adder. So, the total computational time for $g^e \bmod n$ is equal to $6 \cdot k^2 \cdot T$.

---

**Algorithm 1** Montgomery Exponentiation Algorithm

---

**Input:** $g = g_{k-1}, g_{k-2} \cdots g_0$, $e = e_{k-1}, e_{k-2} \cdots e_0$, $e_{k-1} = 1$, and $n = n_{k-1}, n_{k-2} \cdots n_0$. $0 \leq g < n$, $A = 2^k \bmod n$, $e\_2k = 2^{2k} \bmod n$

**Output:** $A = g^e \bmod n$

$g' \leftarrow MontP(g, e\_2k);$ ▷ *MontP* is Montgomery multiplication

**for** $i = k - 1$ downto 0 **do**
$A \leftarrow MontP(A, A)$ ;
**if** $e_i == 1$ **then**
$A \leftarrow MontP(A, g')$;
**end if**
**end for**
$A \leftarrow MontP(A, 1)$ ;
**Return** *A*;

---

Equation 1 states to encrypt a message, we need to get the multiplication result of $g^M$ and $r^n$ exponent pair. Algorithm 2 presents the proposed Montgomery Simultaneous Exponentiation (MSE) to calculate different exponents with two bases, which uses the Left-to-Right method. As shown, four precomputed parameters have to be obtained first. The for loop is performed when the exponents $e_0$ and $e_1$ are not zeros. Thus, the computation time of the MSE depends on the exponents, more ones in the exponents mean more montgomery multiplications. The computation time for the MSE with $k$-bit bases and exponents is given by $(7/4).k.T_{MontP}$, where $T_{MontP}$ is the computational time of a single montgomery multiplication [19].

---

**Algorithm 2** Montgomery Simultaneous Exponentiation (MSE) Algorithm

---

**Input:** $g_0 = g_{0k-1}, g_{0k-2} \cdots g_{00}$, $g_1 = g_{1k-1}, g_{1k-2} \cdots g_{10}$, $e_0 = e_{0k-1}, e_{0k-2} \cdots e_{00}$, $e_{0k-1} = 1$, $e_1 = e_{1k-1}, e_{1k-2} \cdots e_{10}$, $e_{1k-1} = 1$, and $e\_2k = 2^k \bmod n$

**Output:** $A = g_0^{e0} g_1^{e1} \bmod n$

$g_0' \leftarrow MontP(g_0, e\_2k);\ g_1' \leftarrow MontP(g_1, e\_2k);$
$g_{01}' \leftarrow MontP(g_0', g_1');\ \ A \leftarrow MontP(e\_2k, 1);$

**for** $i = k - 1$ downto 0 **do**
$A \leftarrow MontP(A, A);$
**switch** $e_{0i}, e_{1i}$ **do**
  **case 0, 1**
  $A \leftarrow MontP(A, g_0');$
  **case 1, 0**
  $A \leftarrow MontP(A, g_1');$
  **case 1, 1**
  $A \leftarrow MontP(A, g_{01}');$
**end for**
$A \leftarrow MontP(A, 1);$
**Return** $A$

---

Figure 3 represents the proposed architecture of the MSE core, which has three main components: the montgomery multiplication, RAM-based unite for the precomputed parameters, and finite-state-machine control unite. Initially, the four precomputed parameters are calculated and stored. The for loop in Algorithm 2 is applied by scanning the bit-stream of the exponents $e_0$ and $e_1$, then one of the cases is applied accordingly. Note that in each iteration, the value of $A$ is updated so the right result is maintained. The final result is obtained by performing a single montgomery multiplication $MontP(A,1)$, which converts it back to the natural representation.

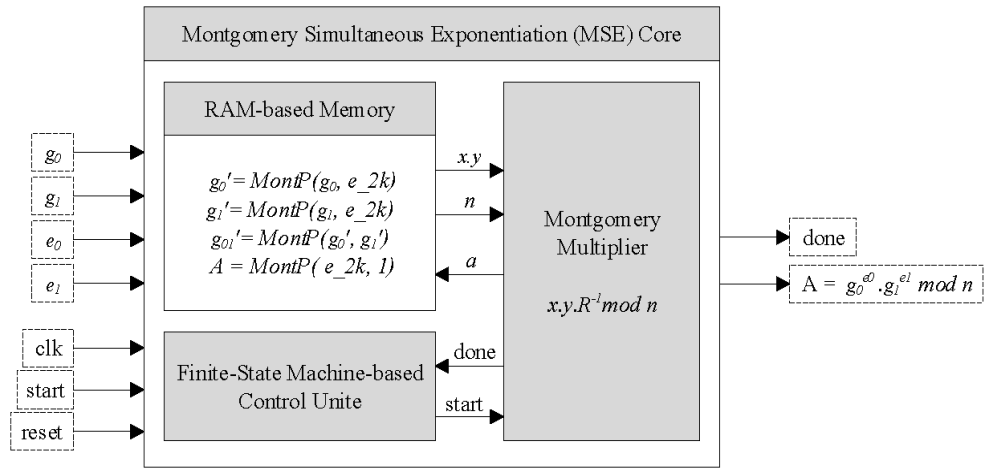

Figure 3. Architecture of the proposed Montgomery Simultaneous Exponentiation (MSE) core.

## 3.2. Domain of the Loaded-encrypted Pixels

In any field $Z_n^*$, all values are $\in [0, n-1]$, and any applied operation over these values must get the result into the field by applying *mod n* operation. In equation 1, the mod is *mod $n^2$*, which means the field is squared, and the results are going to be out of the $\in [0, n^2-1]$ range. The pixels in cover image are gray-scale levels $\in [0, 255]$, and encrypting these pixels will generate results up to ($n^2 - 1$). These encrypted pixels are viewed as a loaded-encrypted stego image, and rendering it is not possible in a gray-scale image. The only way to overcome this rendering issue is to divide each pixel into 3 bytes, each byte represents a channel in RGB image. Figure 4 illustrates the process of rendering the encrypted pixels into RGB stego image. The loaded-encrypted stego image is sent to the receiver to decrypt and extract the secret data out of the RGB stego image.

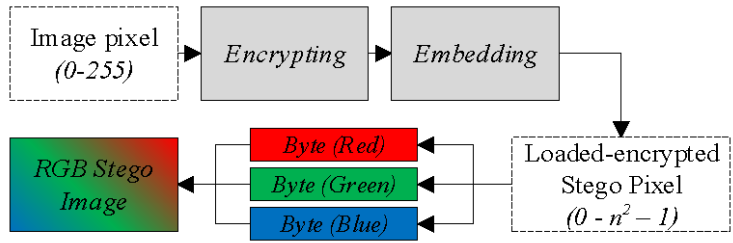

Figure 4. Rendering the loaded-encrypted image into RGB stego image.

## 3.3. Image Decryption and Extraction

On the receiver side, the RGB stego image is entered into a reversible separable decryption and extraction mechanism. The decryption process is applied to every single pixel in the received image. A private key $\lambda$ is required as stated in equation 2. The receiver will be able to extract the secret data without decrypting pixels of the stego image. Decrypting the received image will generate the original cover image using the private key. Using homomorphic property in Equation 3, the pixel P is equal to $DECpr(E_{M1}, E_{M2}) = M_1 + M_2$. Extraction process is done by comparing encrypted pixel pairs $E_{M1}$ and $E_{M2}$. If $E_{M1} > E_{M2}$, the secret bit is 1, and if the $E_{M1} < E_{M2}$, the secret bit is 0. The extraction is done when all encrypted pairs are scanned.

## 4. EXPERIMENTAL RESULTS AND DISCUSSIONS

In this section, the results for the hardware implementation of the steganographic cryptosystem are presented. Verilog Hardware Description Language (HDL) is used to implement the proposed cryptosystem. Verifying the performance of the proposed design is achieved by using the FPGA platform devices which are provided by Xilinx [10], where two FPGA families are used, Artix-7 and Kintex-7 devices [10]. Both FPGA devices are fabricated using the common 28nm technology. The Kintex-7 targets the high-density complex applications such as those in 3G and 4G communications, while the Artix-7 FGPA provides mid-performance for the applications running over power-sensitive systems including the vision cameras and low-end wireless networks. The hardware implementation has been fully synthesized, translated, placed and routed using the new Xilinx Vivado 2018.2 design suite. A balance design strategy is applied as an optimization goal. A time constraint is applied, and the timing constraints report provides a zero timing error. Table 1 shows the results for the proposed steganographic cryptosystem after place and route.

Embedding-encryption module implementation provides a running frequency up to 135.2 MHz using 193 slices of the available resources in Artix-7 FGPA. In Decryption-extraction module implementation, up to 166.7 MHz of running frequency is achieved with 334 slices. Block RAMs (BRAMs) are used as a memory to read the cover image and write the loaded encrypted stego image. More BRAMs are utilized when the size of the cover image is increased. The modules in Kintex-7 provides higher running frequencies between 250 MHz to 333.3 MHz in embedding encryption and decryption-extraction, respectively. The total on-chip power consumption is measured through calculating the power required by the resources in the hardware implementation such as clock, slices (i.e. LUTs and registers) and BRAMs unites at the maximum running frequencies. For the embedding-encryption module, our design consumes 0.139 Watt for 64x64 cover image to 0.134 Watt for 256x256 cover, while the decryption-extraction module takes up to 0.362 Watt of power consumption for 256x256 cover image. The results in Table 1 can tell that our proposed cryptosystem is suitable for the limited-resources low-power embedded systems.

Table 1. Place and route results for encryption and decryption processes.

| Size | Device | Encryption | | | | | Decryption | | | | |
|---|---|---|---|---|---|---|---|---|---|---|---|
| | | Slices | F.F | LUT | BRAM | Freq. | Slices | F.F | LUT | BRAM | Freq. |
| 64 | Artix | 148 | 390 | 409 | 1.5 | 135.2 | 326 | 971 | 768 | 4 | 166.7 |
| 128 | | 151 | 388 | 411 | 5 | 134.8 | 317 | 980 | 774 | 8.5 | 166.5 |
| 256 | | 193 | 409 | 469 | 20 | 133.3 | 334 | 1000 | 819 | 34 | 166.6 |
| 64 | Kintex | 155 | 390 | 429 | 1.5 | 286 | 333 | 967 | 854 | 4 | 333.3 |
| 128 | | 165 | 388 | 430 | 5 | 270.27 | 347 | 980 | 866 | 8.5 | 333.3 |
| 256 | | 179 | 409 | 499 | 20 | 250 | 349 | 1000 | 850 | 34 | 333.3 |

## 4.1. Prototyping the Proposed Image Steganographic Cryptosystem

FPGA device, MATLAB program is used to convert the cover image to a hex format file. The file is loaded into the BRAMs of the FPGA device. Each pixel is read from a specific location (i.e. address) and got it ready for embedding the secret data and encrypt it. The loaded-encrypted pixel is written back to the BRAM at the same location. Once all pixels are loaded by the secret data and encrypted, the loaded-encrypted pixels are extracted from the BRAM and converted to a hex readable file. This file is forwarded to a MATLAB program for the rendering purposes. As mentioned earlier, the rendering process makes the extracted file from the FPGA to be shown as RGB stego image, which is done by allocating a single byte to each channel: red, green and blue. Table 2 represents four gray-scale images (lena, camera man, crowd and barbara) as cover images for the proposed steganographic cryptosystem. Each is represented in 3 different sizes: 64x64, 128x128 and 256x256.

Table 2. Four different sizes gray-scale cover images.

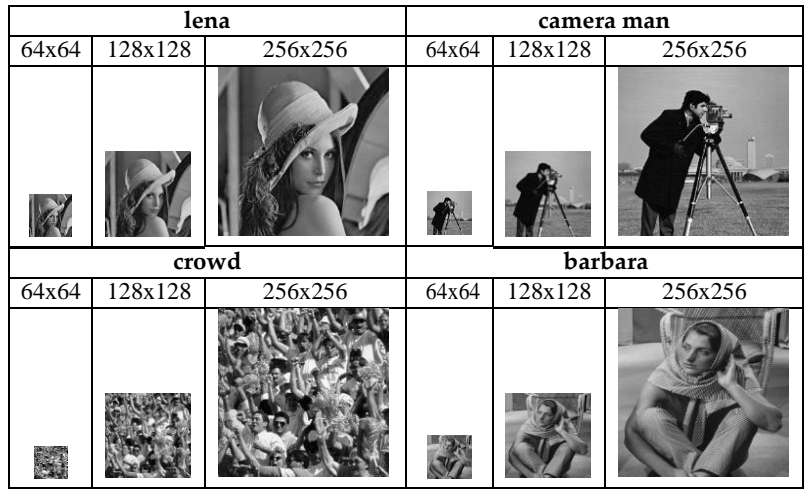

These cover images are the input to the cryptosystem, the secret data as well. For example, if the 64x64 cover image is the input, the secret data has 64x64 or 4096 bits are going to be concealed. Table 3 shows RGB stego images for the cover images under different sizes. Note that for each original cover image at any size, a random secret data is generated to achieve randomness in the results (i.e. RGB stego images). As shown, the stego images are rendered differently from size to another. The stego images are ready for transmitting to the receiver for recovering original images and extracting secret data reversibly.

Table 3. RGB stego images for cover images.

| lena ||| camera man |||
|---|---|---|---|---|---|
| 64x64 | 128x128 | 256x256 | 64x64 | 128x128 | 256x256 |
| 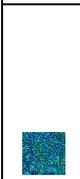 | 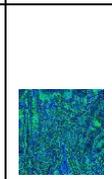 | 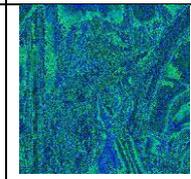 | 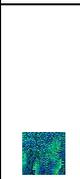 | 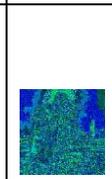 | 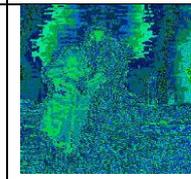 |
| crowd ||| barbara |||
| 64x64 | 128x128 | 256x256 | 64x64 | 128x128 | 256x256 |
| 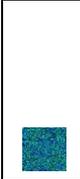 | 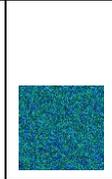 | 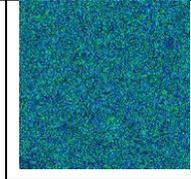 | 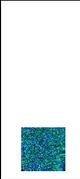 | 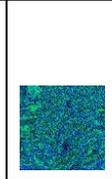 | 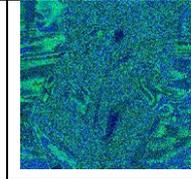 |

The receiver will perform the reversible decryption and extraction processes to recover the original cover image and get the secret data. To perform that, the received stego image is converted to a text hex format file by running a MATLAB program. Then, this file is loaded to the BRAMs of the FPGA, where each pixel is decrypted, and bits of secret data are extracted. The decrypted pixel is written back to the BRAM at same address of the encrypted pixel. This is applied to all pixels in the stego image. Once all pixels are decrypted, a file is extracted from the BRAMs in a hex readable format. The file is forwarded to a MATLAB program to gather all decrypted pixel and render them into a gray-scale original cover image. Table 4 represents the decrypted original cover images for the RGB stego images at the receiver side.

Table 4. Decrypted cover images at the receiver side.

| lena ||| camera man |||
|---|---|---|---|---|---|
| 64x64 | 128x128 | 256x256 | 64x64 | 128x128 | 256x256 |
| 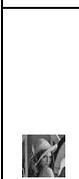 | 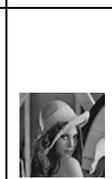 | 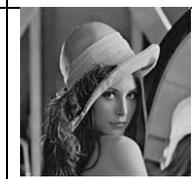 | 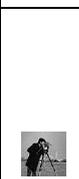 | 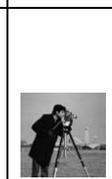 | 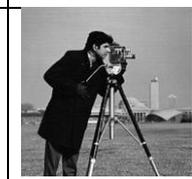 |
| crowd ||| barbara |||
| 64x64 | 128x128 | 256x256 | 64x64 | 128x128 | 256x256 |
| 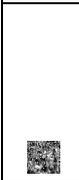 | 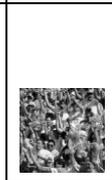 | 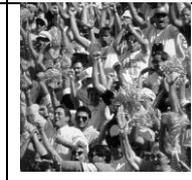 | 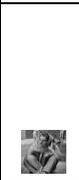 | 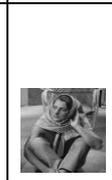 | 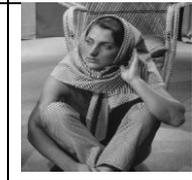 |

Peak Signal-to-Noise Ratio (PSNR) is a subjective indicator for the quality of an image, which is measured in decibels (dB). The higher PSNR value indicates maintaining better quality. In the proposed steganographic, 100% PSNR is achieved since there is no loss in cover images, thus the receiver can recover any kind of cover images perfectly. For embedding rate, the proposed cryptosystem always offers a fixed rate of 1 bpp. As quantitative metric for the performance, the throughput of the steganographic cryptosystem is defined as the total bits that cryptosystem can process per time unit. The following equation shows how to calculate the throughput for the proposed cryptosystem.

$$\text{Throughput} = \frac{\text{Input-Data} \cdot \text{Frequency}}{\text{Clock Cycles of Encryption or Decryption}} \qquad (7)$$

Table 5 shows the performance results in terms throughput and Frame Per Second (fps) for the embedding-encryption and decryption-extraction hardware FPGA implementations in different cover images. It's expected to get large number of clock cycles in the proposed cryptosystem due to the intensive field operations in the paillier cryptosystem. To encrypt a single pixel, 1520 clock cycles are required, and decrypt it requires 1460 clock cycles. Note that the decryption is performed over RGB pixels. From Table 5, FPS rate is calculated as number of pixels in one image divided by the throughput. For instance, it is possible to process 22.3 fps in embedding-encryption module, and 28.6 fps in decryption-extraction module for 128x128 cover image in real-time embedded system. For 1-minute period, the proposed cryptosystem can process a video of 21.9 MB size at 22.3 fps and 84.3 MB size at 28.6 fps in embedding-encryption and decryption extraction module hardware implementations, respectively.

Table 5. Performance results for different cover images.

| Cover Image | Size | Throughput (Kpps) | | FPS | |
|---|---|---|---|---|---|
| | | Encryption | Decryption | Encryption | Decryption |
| lena | 64 | 192.7 | 207.6 | 47.1 | 50.7 |
| Cameraman | | | | | |
| crowd | | | | | |
| barbara | | | | | |
| lena | 128 | 365.2 | 468.5 | 22.3 | 28.6 |
| Cameraman | | | | | |
| crowd | | | | | |
| barbara | | | | | |
| lena | 256 | 676.4 | 937.1 | 10.4 | 14.3 |
| Cameraman | | | | | |
| crowd | | | | | |
| barbara | | | | | |

Table 6 represents a comparison between the proposed hardware implementation of the paillier cryptosystem and other implementations. Our design outperforms the works in terms of PSNR, utilized slices and frequency. In paper [6], the hardware implementation is designed using interpolation expansion method. The design achieved a high performance in terms of the throughput, which is capable of processing about 7640 pixels in a single second using Virtex-6 FPGA Device. On the other hand, a huge amount of resources as register and LUTs are utilized, where it requires around 10 times of the resource utilized by our design, and our design consumes less power by 91.5% than the work in [6]. In [20], a pipelined reversible hardware architecture for secret water marking embedding is proposed, where the Reversible Contrast Mapping (RCM) algorithm is adopted for embedding and extraction secret data. The architecture offers low-complexity fast processing through breaking the critical path of the design into 6 pipelined stages. However, our hardware architecture is faster, requires less resources and achieves higher throughput than the work in [20].

The work in [21] represents methods of encapsulating secret data into a group *n* of pixels in a (2*n*+1)-ary notational system. The hardware implementation offers high fps but with low PSNR and high utilized resources, where 16% of the total resources in Virtex -7 FPGA device is roughly 12,000 logic cells. The power consumption in [21] is very high, it is 3.807 Watt, while our hardware implementation consumes between 0.139 Watt to 0.362 Watt of power. We should mention that the hardware implementations in [6], [7], [20] and [21] do not apply any kind of cryptographic algorithms. To the best of our knowledge, our proposed reconfigurable implementation for image steganography using paillier cryptosystem is the first to present in the literature. Figure 5 represents a comparison between our proposed steganographic scheme with other works [22] and [23] in terms of image quality and embedding rate. The image quality is always 100% PSNR regardless of the embedding rates.

Table 6. Performance comparison with existing implementations for image steganography.

| Ref. | Image Size | Device | bpp | PSNR | Slices | LUT | F.F | Freq. | Throughput (Kpps) |
|---|---|---|---|---|---|---|---|---|---|
| [6] | 328x264 | Virtex-5 | 1 | x | x | 9,715 | 4883 | 96.4 | 7340.17 |
|  |  | Virtex-6 | 1 | x | x | 8,564 | 4908 | 100.4 | 7640.63 |
| [7] | 128x128 | Spartan-6 | 1 | 51.1444 | 160 | 322 | 214 | 104.971 | x |
| [21] | 512x512 | Virtex-7 | 1 | 45.11 | 16% | 1% | 2% | 144 | x |
| [20] | 32x32 | Spartan-3 | 0.46 | 29.46 | 9881 | 11291 | 9347 | 98 | 100.3 |
| **Ours** | 128x128 | Artix-7 | 1 | 100% | 151 | 774 | 980 | 134.8 | 370 |
|  | 256x256 | Kintex-7 | 1 | 100% | 179 | 850 | 1000 | 250 | 680 |

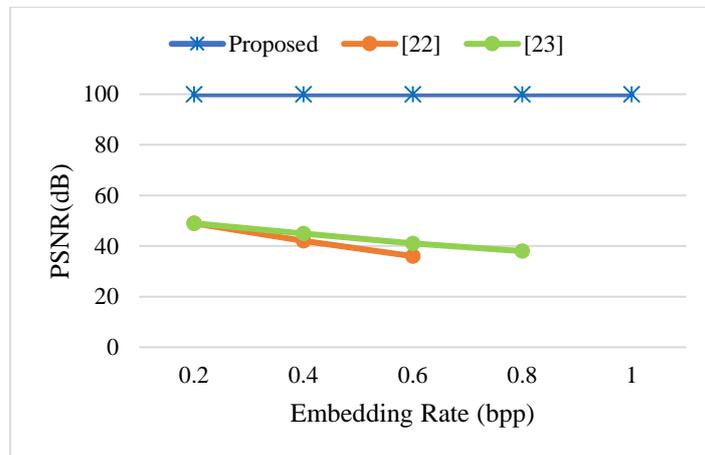

Figure 5. Comparison of PSNR vs embedding rate performance between the proposed steganographic scheme and others.

## 5. CONCLUSIONS

In this paper, a hardware architecture for a separable image steganographic scheme using paillier cryptosystem is designed and implemented on reconfigurable FPGA platforms. An efficient montgomery simultaneous exponentiation is implemented for performing exponentiation operations in encryption and decryption of paillier cryptosystem. The place and routed results have demonstrated a high performance in terms of speed, utilized resources and power consumption in the new Kintex-7 and Artix-7 Xilinx FPGA devices. Our proposed steganographic achieves a guaranteed 100% PSNR for any kind of cover images with a fixed 1 bpp of embedding capacity. In real-time embedded system, our proposed hardware architecture is able to embed and encrypt up to 47.1 fps of cover images. In decryption-extraction processes, the hardware architecture can perform up to 28.6 fps of RGB stego images. The proposed secure architecture offers a small footprint which can be utilized in many embedded applications as graphic processing engines and accelerators.

## ACKNOWLEDGEMENTS

This work was supported in part by the Natural Sciences and Engineering Research Council (NSERC) of Canada and in part by the Regroupement Strategique en Microelectronique du Quebec (ReSMiQ).